\documentclass[twocolumn]{aastex631}  

\received{March 7, 2024}
\revised{April 20, 2024}
\accepted{April 23, 2024}

\shorttitle{Glints from Space Debris in LSST}
\shortauthors{Tyson et al.}

\begin{document}

\title{Expected Impact of Glints from Space Debris in the LSST}


\correspondingauthor{J. Anthony Tyson}
\email{tyson@physics.ucdavis.edu}

\author[0000-0002-9242-8797]{J. Anthony Tyson}
\altaffiliation{IAU Centre for the Protection of the Dark and Quiet Sky from Satellite Constellation Interference}
\affiliation{Department of Physics \& Astronomy, University of California, Davis, CA, USA}

\author[0000-0002-2343-0949]{Adam Snyder}
\affiliation{Department of Physics \& Astronomy, University of California, Davis, CA, USA}

\author[0000-0001-7445-4724]{Daniel Polin}
\affiliation{Department of Physics \& Astronomy, University of California, Davis, CA, USA}

\author[0000-0003-1305-7308]{Meredith L. Rawls}
\altaffiliation{IAU Centre for the Protection of the Dark and Quiet Sky from Satellite Constellation Interference}
\affiliation{Department of Astronomy/DiRAC/Vera C. Rubin Observatory, University of Washington, Seattle, WA, USA}

\author[0000-0001-5250-2633]{\v{Z}eljko Ivezi\'{c}}
\affiliation{Department of Astronomy/DiRAC/Vera C. Rubin Observatory, University of Washington, Seattle, WA, USA}

\begin{abstract}

We examine the simple model put forth in a recent note by
Loeb regarding the brightness of space debris in the size range of 1--10 cm and their impact on the Rubin Observatory Legacy Survey of Space and Time (LSST) transient object searches. Their main conclusion was that ``image contamination by untracked space debris might pose a bigger challenge [than large commercial satellite constellations in LEO]". 
Following corrections and improvements to this model, we calculate the apparent brightness of tumbling low-Earth orbit (LEO) debris of various sizes, and we briefly discuss the likely impact and potential mitigations of glints from space debris in LSST. \added{We find the majority of the difference in predicted signal-to-noise ratio (S/N), about a factor of 6, arises from the defocus of LEO objects due to the large Simonyi Survey Telescope primary mirror and finite range of the debris.} The largest change from the
Loeb
estimates is that 1--10 cm debris in LEO pose no threat to LSST transient object alert generation because their S/N for detection will be much lower than estimated by
Loeb due to defocus. We find that only tumbling LEO debris larger than 10 cm or with significantly greater reflectivity, which give 1 ms glints, might be detected with high confidence (S/N$>$5). We estimate that only one in five LSST exposures low on the sky during twilight might be affected. More slowly tumbling objects of larger size can give flares in brightness that are easily detected; however, these will not be cataloged by the LSST Science Pipelines 
because of the resulting long streak.

\end{abstract}


\keywords{Transient sources (1851); Light pollution (2318); Sky surveys (1464); Ground-based astronomy (686); Artiﬁcial satellites (68)}

\section{Introduction}

Low-Earth orbit (LEO) debris larger than 10 cm diameter are tracked and cataloged by the US Space Surveillance Network, available via \texttt{space-track.org}.
Debris smaller than this size, which is the subject of \citet{loeb24}, are a potential issue because their orbits remain unknown, even if historical follow-up of an astrophysical candidate is pursued. Therefore, for these objects the origin of short, bright instances of specular reflection (glints) seen by telescopes on Earth cannot readily be distinguished as astrophysical or artificial in nature. \added{From radar data, there is a log number--log size power-law distribution of space debris in LEO with slope $\sim -1.8$ \citep{arnold2023radar, tarran2021, krisko2014new, liou2001new} over the size range 1--10 cm. This may present challenges to the Rubin Observatory Legacy Survey of Space and Time (LSST) search for faint transient astrophysical objects.} This was the main thesis of \citet{loeb24}, and we agree this is a potential concern for certain debris geometries. 

In this note, we examine and update two assumptions made in \citet{loeb24} and describe two mitigations that lessen the impact. First, \citet{loeb24} assumes incorrectly that the debris is in focus. \added{Because of the large primary mirror (8.4 m), an arbitrarily small object observed at a finite range by the LSST Camera will appear partially out of focus; this defocusing effect will cause a reduction of the peak surface brightness of the object, as described in detail in \citet{bektesevic18}}. Second, tumbling debris that could potentially confuse LSST transient detection and classification will not remain stationary to the LSST field of view. Instead, it will produce a series of slightly trailed glints as the image of the debris object passes across the focal plane. This means that glints may be recognized morphologically, and the resulting alerts may be classified accordingly.



\section{Defocus and Apparent Brightness}

\citet{loeb24} claims that LEO debris of 1 cm size range may interfere with LSST transient searches. Small debris objects capture a fraction of the Sun's flux, relaying it to the telescope at a slant range of about 1000 km. If the object is tumbling, the Sun's relayed flux illuminates the telescope for a small fraction of the exposure time, diluting its contribution. We first estimate the brightness of a 1 cm radius piece of debris with albedo 0.1 orbiting at 500 km that produces a glint of 1 ms, as observed by LSSTCam at $60^\circ$ angle from zenith. This assumes a fraction of incident sunlight (0.1) is diffusely reflected uniformly from the debris over one hemisphere, as described by \citet[Equation (2)]{loeb24}. The total flux from the glint will scale linearly with this fraction. In Section \ref{sec:discuss}, we also simulate and discuss the high albedo case of specular reflection. 

The Rubin Observatory simulation software tools for modeling the LSSTCam instrument throughput\footnote{\url{https://github.com/lsst/rubin_sim}} sets the AB magnitude of the Sun in the 500 nm $g$-band at $-26.5$. This results in a total flux from the glint equal to that of a 16.6 $g$-band mag object observed for 1 ms (or equivalently, a 27.7 $g$-band mag object observed for 30 s). This is well below the LSST $g$-band limiting magnitude of 24.4 \citep{bianco2022} and 0.4 mag dimmer than the 16.2 mag found by \cite{loeb24} in their Equation 2 (this change is likely caused by different assumptions regarding the AB magnitude of the Sun in $g$-band as measured by LSSTCam). Therefore a 10 cm object with these properties will yield total glint flux equal to a 11.6 $g$-band mag object observed for 1 ms (or equivalently, a 22.7 $g$-band mag object observed for 30 s).

\begin{figure*}[ht!]  
\plotone{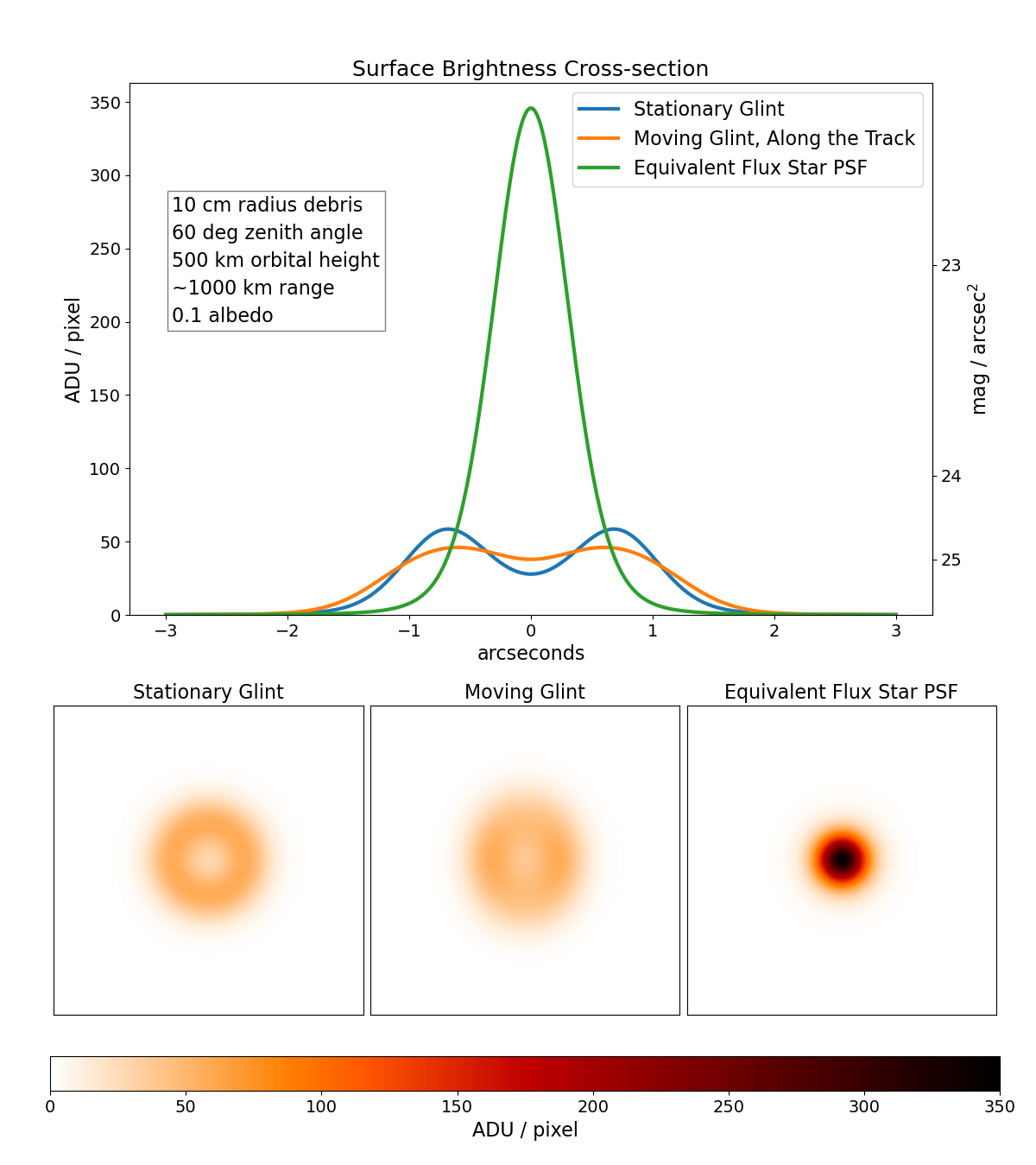}
\caption{Effect of defocus of small LEO objects (10 cm radius debris) due to the large LSST primary mirror. {\it Top}: the simulated surface-brightness cross section of a faint stationary 1 ms glint in a 30 s LSST exposure in blue, an in-focus star of the same net flux (22.7 mag) with PSF size of $0.''73$ in green, and the same glint simulated with the added effect of moving across the focal plane in orange, assuming a 500 km altitude orbit. {\it Bottom left}: the simulated surface brightness profile of the faint stationary 1 ms glint in a 30 s LSST exposure. {\it Bottom center}: the same glint simulated with the added effect of moving across the focal plane. {\it Bottom right}: the in-focus star with PSF size of $0.''73$. In this simulation, 1 ADU = 1 electron. The actual LSST matched-filter source-detection process, the PSF convolved with these images, produces the same conclusion. This simulated glint debris assumes an albedo of 0.1; the peak brightness scales linearly with albedo.
\label{fig:surface_brightness_profiles}}
\vspace{20em}  
\end{figure*}

Next, we consider the effect of defocusing on the peak brightness of the glint. Using LEO satellite simulation software\footnote{\url{https://github.com/Snyder005/leosim}}, \added{which was developed to simulate both satellite streaks and glints and performs well compared to observations from a smaller telescope,} we calculate the surface-brightness profile of a glint from a 10 cm radius debris object orbiting at 500 km that is observed at $60^\circ$ zenith angle by Rubin Observatory (1000 km range). The debris is modeled as a uniform 10 cm radius disk convolved with a Kolmogorov atmosphere kernel with $0.''73$ seeing---typical for the Rubin Observatory site---and a defocus kernel (derived from the primary and secondary mirror geometries) to produce a stationary image of the glint. An object at this orbit will move $1-2''$ relative to the LSST field of view during a 1 ms glint, leaving a short linear trail. A moving image of the glint is calculated by convolving its stationary image with a $1''$ width boxcar window. This width is chosen by considering both the debris tangential orbital velocity and the glint duration. Figure \ref{fig:surface_brightness_profiles} shows the surface-brightness profiles of both the stationary and moving glint from the 10 cm radius object, as well as an in-focus star of 22.7 mag with point-spread function (PSF) size of $0.''73$ that has an equivalent total flux. Also included are the corresponding surface-brightness profile cross sections. The peak surface brightness of the stationary and moving glint images is lower by a factor of 5.9 compared to the PSF of a star with the same net flux ($0.''73$ FWHM). The effect is more pronounced for lower orbits as debris objects become more out of focus. This simulation also shows that there is a relatively minor decrease in peak surface brightness due to the motion of the already-defocused debris relative to the focal plane.

We next examine the sensitivity of LSST to these defocused, moving glints. \citet{loeb24} quotes a value for LSST sensitivity of 14 mag, derived by scaling a 25 mag limit to the 1 ms effective exposure time of a glint compared to the 30 s image exposure time. However, both defocusing and motion distribute the total glint flux over more pixels, resulting in a smaller signal per pixel. Transient detection thresholds are based on a peak in the PSF-convolved image (maximum likelihood flux estimate), and therefore both effects result in a lower signal-to-noise ratio (S/N) for detection. For any detection process, there is an accuracy$-$completeness relation; a lower S/N therefore results in detection incompleteness and lower alert efficiency. Taken together, we predict about a factor of 6 fewer glint detections due to the difference in predicted S/N than assumed in \citet{loeb24}.

The 30 s exposure $g$-band 24.4 mag \citep{bianco2022} 
sensitivity limit scales to a 13.2 mag limit for a 1 ms glint. For LEO debris in a 500 km altitude orbit at a range of approximately 1000 km, the defocus effect will decrease the LSST brightness sensitivity to the debris by the same factor as the peak surface brightness shown in Figure \ref{fig:surface_brightness_profiles}. This results in an effective debris sensitivity limit of 11.3 mag. Therefore, we conclude a 10 cm radius object with an 11.6 $g$-band mag glint will be at the edge of the LSST detection limit. 

By processing the simulated glint postage stamp images using the LSST Science Pipelines \citep{bosch19} source-detection task, which convolves the image with a Gaussian sized to the image PSF prior to detecting sources, we verify that glints from debris under the above assumptions are not detected using the default detection parameters. We note this assumes the debris uniformly reflects one-tenth of the incident sunlight, which will not be the case for glints from surfaces designed to be highly specularly reflective. Therefore, we expect only a small fraction of 10 cm size debris to be detectable in LSST, mainly those with specular, mirror-like surfaces (see Section \ref{sec:discuss}). In general, however, the large population of LEO debris below a few centimeters in size may pose little challenge for LSST transient science.



\section{Morphological Signature of Tumbling LEO Debris and Mitigations}

Consider the example in \citet{loeb24} of a piece of flat, reflective debris in LEO tumbling once per second. Due to the angular size of the Sun, this creates faint glints of $\sim1$ ms duration. These have been seen in Zwicky Transient Facility (ZTF) data \citep{karpov22}. Given the angular size of the Sun and the duration of visible glints, the object would have to be spinning rapidly enough to generate a beaded line of glints repeating every second, rather than a single glint, during the $5-10$ s it takes to cross the LSSTCam focal plane. The LSST Science Pipelines should be able to identify this unique signature of spatially repeated flashes to prevent downstream LSST transient classifiers from triggering on glints from debris and contaminating transient searches.

Finally, we note the presence of numerous glints may enhance the utility of double exposures during each LSST visit to a sky position. Early estimates gave some impetus to the ``snap'' approach---two 15 s exposures per 30 s visit to a sky position---which is the baseline LSST plan, with the exception of the $u$ band, which will have a single 30 s exposure per visit. Two snaps, or difference with sky template, will allow for improved morphological detection of glints in individual exposures as they will not be in both images and could more readily be removed from alerts and catalogs.  As discussed below, the known power-law distribution of LEO debris size allows an estimate of the number of these faint glints in LSST data.  

\section{Discussion} \label{sec:discuss}

While glints from 1 to 5 cm radius debris at LEO are sufficiently faint to escape LSST detection and alerts, with the notable exception of sky brightness effects \citep{kocifaj21}, this is not true for debris 10 cm and larger. For orbits below 600 km, as with bright satellites, such bright debris glints are most common within 2 hr of twilight and at low sky elevations. With approximately $10^4$ such objects \citep{arnold2023radar}, how many will be seen in the survey? To estimate the number seen by LSST, we need to fold in LSST operations.  Conveniently, \citet{hu2022satellite} have done a full operations simulation for a constellation of about 40,000 LEO satellites. They found roughly 10\% of 30 s visits have at least 1--2 satellite streaks, depending on telescope ALT (angle above horizon). We have also used this same observing database to examine the cumulative number of satellites observed versus solar angle. Given that half of LSST planned observations below $55^\circ$ ALT\footnote{Lynne Jones, private communication}, and using the same satellite population assumptions from \citet{hu2022satellite}, we find that there may be as many as one glint per five LSSTCam observations at ALT $< 50.^\circ$

If debris tumbles once per second, this would be repeated several times along the trail in the LSSTCam focal plane. Due to their morphology, it should usually be possible to distinguish chains of glints from real astrophysical sources. The assumption of 1 ms glints in \citet{loeb24} was based on observed chains of glints separated by $\sim 1$ s in ZTF images \citep{karpov22}. An average albedo of 0.1 is meant to cover a range of debris. In reality, most centimeter-scale debris are expected to be dominated by specular flat surfaces with albedos approaching 1, tumbling on a range of timescales \citep{fankhauser23}. Thus a broad distribution of glint duration is expected, extending up to second-timescale flares \citep{vestrand22}. Unless replenished by collision debris, the population of fast tumbling small debris can be depleted over time by atmospheric drag and eddy current braking in the Earth's magnetic field; low-mass fast-spinning debris of planar geometry are depleted first.

It is clear we are presently witnessing an exponential increase in the number of LEO satellites and that this will continue well into the 2030s, coincident with LSST Operations. In their normal operational orbits, most larger satellites will not produce millisecond faint glints. Rather, they typically reflect sunlight for 1 s or more and produce streaks or flares that span hundreds of arcseconds across the sky. As the number of LEO satellites increases above 50,000, there is an increased likelihood of multiple collisions producing small spinning debris \citep{long2020impacts}. Since there are well over 100,000 credible constellation satellites planned \citep{mcdowell2023}, this is a real concern. Indeed there are currently over $10^6$ proposed LEO satellites in applications to the International Telecommunication Union (ITU) \citep{falle2023one}.

Although \citet{loeb24} did not consider high-LEO debris, there is some debris at 1000 km orbits and may be more in the future.  We have repeated our analysis for 1--10 cm debris orbiting at 1000 km and observed at $60^\circ$ zenith angle, resulting in a range of 1750 km. The debris object moves more slowly across the focal plane and is slightly more in focus, but is also fainter. We find that overall, this decreases the peak brightness of the glint image by a factor of 1.4 and lowers the sensitivity by 0.4 mag.

LEO satellite brightness, including flares in brightness, is possible to model accurately given full knowledge of the bidirectional reflectance distribution function (BRDF) of all satellite components \citep{fankhauser23}. However, we do not have this capability for space debris. Radar surveys have detected LEO debris of $\sim 5$ cm size with rapid variations at ranges near 1000 km \citep{muntoni21}, and modeling of debris is more likely to be successful with a focus on glint rates or patterns rather than brightnesses. We point out that the model of a uniformly diffuse debris object with an albedo maximum of 1 can be incorrect for mirror-like surfaces \added{such as shards of multilayer insulation, and even small debris objects will often have a combination of diffuse and reflective surfaces}. The BRDF of specular surfaces can have a sharp peak reaching 100 over $1^\circ$ angular range
\citep{fankhauser23}. Conveniently for LSST, lab measurements show a significant minimum at LSST filter wavelengths for various spacecraft materials \citep{bowers2011broadband}.

The main LSST program for short, faint time-domain probes is the ``Deep Drilling Fields'' (DDF). In contrast to the wide-fast-deep main survey, the DDF program involves observing a well-studied field for about 1 hr per night with continuous 15 s exposures. Any satellite glinting in the DDF data will have broad time coverage and will appear in multiple images or multiple times in one image and will provide an opportunity to better characterize the LEO debris population and understand its impacts on LSST science and beyond. 

In the future, next-generation CMOS detectors will allow rapid readout, stereoscopic optical surveys of the 0.1 s glints from debris in LEO \citep{vestrand22}.
The largest remaining uncertainty is the actual LEO debris glint rate distribution. In addition, concerns exist for the proliferation of the smallest debris yielding a change in the sky background brightness level, which we do not describe in detail here (see \citealt{kocifaj21} for more). Even before the LSST begins in 2025, there is some chance that we will learn more about 16 mag and brighter glints on one second timescales from the Argus Pathfinder \citep{corbett22}. 

\begin{acknowledgments}

We thank Eric Bellm, Jim Bosch, Forrest Fankhauser, Craig Lage, Leanne Guy, Lynne Jones, Robert Lupton, Tom Vestrand, and Peter Yoachim for helpful comments. We also acknowledge helpful comments by an anonymous referee. A.S. and J.A.T. acknowledge support from NSF grant AST-2205095. This paper makes use of LSST Science Pipelines software developed by Vera C. Rubin Observatory. This code is available as free software at \url{https://pipelines.lsst.io}.

\end{acknowledgments}


\vspace{5mm}


\software{astropy \citep{astropy2022}, LSST Science Pipelines \citep{bosch18,bosch19}
      }

\bibliography{space-debris-glints}{}
\bibliographystyle{aasjournal}

\end{document}